\documentclass[twoside,reqno]{lt7-proc}
\usepackage{epsfig,cite}
\usepackage{amssymb,amsmath}
\usepackage{times}
\begin{document}
\sloppy \raggedbottom
\setcounter{page}{1}

\newpage
\setcounter{figure}{0}
\setcounter{equation}{0}
\setcounter{footnote}{0}
\setcounter{table}{0}
\setcounter{section}{0}



\title{``Gauge'' in General Relativity :\\
  {\small --- Second-order general relativistic gauge-invariant
  perturbation theory ---}
}

\runningheads{Nakamura}{``Gauge'' in General Relativity
}

\begin{start}


\author{Kouji Nakamura}{1},

\address{
    Department of Astronomical Science,
    the Graduate University for Advanced Studies,
    Mitaka 181-8588, Japan
  }{1}


\begin{Abstract}
As in the case of the other gauge field theories, there is so
called ``gauge'' also in general relativity.
This ``gauge'' is unphysical degree of freedom.
There are two kinds of ``gauges'' in general relativity.
These are called the first- and the second-kind of gauges,
respectively.
The gauge of the first kind is just coordinate system on a
single manifold.
On the other hand, the gauge of the second kind arises in the
general relativistic perturbations.
Through the precise distinction of these two concepts of
``gauges'', we develop second-order gauge-invariant general
relativistic perturbation theory.
\end{Abstract}
\end{start}


\section{Introduction}


General relativity is regarded as a gauge theory.
In 1956, Uchiyama pointed out that gravitational field is
introduced by the invariance of the action under local Lorentz
transformations\cite{Uchiyama1956}.
This is due to general covariance in general relativity.
Since this local Lorentz group is a group of coordinate
transformations, coordinate system is called ``gauge'' in
general relativity from this history.


General covariance intuitively states that there is no preferred
coordinate system in nature and it also introduce ``gauge'' in
the theory.
This ``gauge'' is the unphysical degree of freedom and we have
to fix the ``gauge'' or to extract some invariant quantities to
obtain physical results.
Thus, treatments of ``gauge'' are crucial in general
relativity.
This situation becomes more delicate in general relativistic
perturbation theory.


On the other hand, the general relativistic higher-order
perturbation theory is a topical subject in recent physics, for
example, cosmological perturbations, perturbations of a black
hole and a star.
So, it is necessary to formulate the higher-order general
relativistic perturbation theory form general point of view.


In this article, we clarify the notion of ``gauges'' in general
relativity, which is necessary to develop general relativistic
higher-order ``gauge-invariant'' perturbation theory.
The details of this perturbation theory can be seen in
Ref.\cite{KNs}.


\section{``Gauge'' in general relativity}


In 1964, Sachs\cite{Sachs1964} pointed out that there are two
kinds of ``gauges'' in general relativity which are closely
related to the general covariance.
He called these two ``gauges'' as the first- and the second-kind
of gauges, respectively.


{\it The first kind gauge} is a coordinate system on a single
manifold ${\cal M}$.
On a manifold, we can always introduce a coordinate system as a
diffeomorphism $\psi_{\alpha}$ from an open set
$O_{\alpha}\subset {\cal M}$ to an open set
$\psi_{\alpha}(O_{\alpha})\subset {\mathbb R}^{n}$ 
($n=\dim{\cal M}$).
This diffeomorphism $\psi_{\alpha}$, i.e., coordinate system of
the open set $O_{\alpha}$ is called {\it gauge choice} (of the
first kind).
If we consider another open set in $O_{\beta}\subset {\cal M}$,
we have another gauge choice $\psi_{\beta}$ :
$O_{\beta}\mapsto\psi_{\alpha}(O_{\alpha})\subset 
{\mathbb R}^{n}$ for $O_{\beta}$.
The diffeomorphism $\psi_{\beta}\circ\psi_{\alpha}^{-1}$ is
called {\it gauge transformation} (of the first kind), which is
a coordinate transformation : 
$\psi_{\alpha}(O_{\alpha}\cap O_{\beta})$ $\subset$ 
${\mathbb R}^{n}$ $\mapsto$ 
$\psi_{\beta}(O_{\alpha}\cap O_{\beta})$ $\subset$ ${\mathbb R}^{n}$.


According to the theory of a manifold, coordinate systems are
not on a manifold itself but we can always introduce the
coordinate system through a map from an open set in the manifold
${\cal M}$ to an open set of ${\mathbb R}^{n}$.
For this reason, general covariance in general relativity is
automatically included in the premise that our spacetime is
regarded as a single manifold.
The first kind gauge arises due to this general covariance but
it is harmless if we apply the covariant theory on the manifold
in many cases.


{\it The second kind gauge} appears in general relativistic
perturbations.
To explain this, we have to remind what we are doing in
perturbation theory.
First, in any perturbation theories, we always treat two
spacetime manifolds.
One is the physical spacetime ${\cal M}$, which we want to
describe by perturbations, and the other is the background
spacetime ${\cal M}_{0}$, which is prepared for perturbative
analyses by us.
Note that these two spacetimes ${\cal M}$ and ${\cal M}_{0}$
are distinct.
Second, in any perturbation theories, we always write equations
in the form
\begin{equation}
  \label{eq:variable-symbolic-perturbation}
  Q(``p\mbox{''}) = Q_{0}(p) + \delta Q(p)
\end{equation}
as the perturbation of the variable $Q$.
Keeping in our mind that we always treat two different
spacetimes, ${\cal M}$ and ${\cal M}_{0}$, in perturbation
theory, Eq.~(\ref{eq:variable-symbolic-perturbation}) is a
rather curious equation because the variable on the left-hand
side of Eq.~(\ref{eq:variable-symbolic-perturbation}) is a
variable on the physical spacetime ${\cal M}$, while the
variables on the right-hand side of
Eq.~(\ref{eq:variable-symbolic-perturbation}) are variables on
the background spacetime, ${\cal M}_{0}$.
In short, Eq.~(\ref{eq:variable-symbolic-perturbation}) gives a
relation between variables on two different manifolds.


We note that, through
Eq.~(\ref{eq:variable-symbolic-perturbation}), we have
implicitly identified points in these two different
manifolds. More specifically, the point $``p\mbox{''}$ in
$Q(``p\mbox{''})$ on the left-hand side of
Eq.~(\ref{eq:variable-symbolic-perturbation}) is on ${\cal M}$.
Similarly, the point $p$ in $Q_{0}(p)$ or $\delta Q(p)$ on the
right-hand side of Eq.~(\ref{eq:variable-symbolic-perturbation})
is on ${\cal M}_{0}$.
Because Eq.~(\ref{eq:variable-symbolic-perturbation}) is
regarded as an field equation, it implicitly states that the
points $``p\mbox{''}\in{\cal M}$ and $p\in{\cal M}_{0}$ are
same.
This implies that we are assuming the existence of a map 
${\cal M}_{0}\rightarrow{\cal M}$ $:$ $p\in{\cal M}_{0}\mapsto
``p\mbox{''}\in{\cal M}$, which is a {\it gauge choice} (of the
second kind)\cite{J.M.Stewart-M.Walker11974}.


It is important to note that the second kind gauge choice
between ${\cal M}_{0}$ and ${\cal M}$ is not unique to the
theory with general covariance.
Rather, Eq.~(\ref{eq:variable-symbolic-perturbation}) involves
the degree of freedom in the choice of the map ${\cal X}$ $:$
${\cal M}_{0}\mapsto{\cal M}$.
This is called the {\it gauge degree of freedom} (of the second
kind).
This gauge degree of freedom always exists in perturbations of a
theory with general covariance.
If general covariance is not imposed, there is a preferred
coordinate system in the theory, and we naturally introduce this
coordinate system onto both ${\cal M}_{0}$ and ${\cal M}$.
Then, we can choose the identification map ${\cal X}$ using this
preferred coordinate system.
However, general covariance states that there is no such
coordinate system, and we have no guiding principle to choose
the identification map ${\cal X}$.
Indeed, we may identify $``p\mbox{''}\in{\cal M}$ with 
$q\in{\cal M}_{0}$ ($q\neq p$) instead of $p\in{\cal M}_{0}$ by
the {\it different gauge choice} ${\cal Y}$ of the second kind.


\section{Gauge transformations and gauge invariant variables}


To define the perturbation of an arbitrary tensor field $Q$, we
consider the one-parameter family of spacetimes 
${\cal M}_{\lambda}$ so that ${\cal M}_{\lambda}={\cal M}$ and
${\cal M}_{\lambda=0}={\cal M}_{0}$, where $\lambda$ is an
infinitesimal parameter for perturbations.
We regard $\{{\cal M}_{\lambda}|\lambda\in{\mathbb R}\}$ as an
extended manifold.
The gauge choice is made by assigning a diffeomorphism 
${\cal X}_{\lambda}$ $:$ ${\cal M}_{0}$ $\rightarrow$ 
${\cal M}_{\lambda}$ on this extended manifold.
A tensor field $Q$ on ${\cal M}_{\lambda}$ is pulled-back by
${\cal X}_{\lambda}^{*}$ : $Q$ $\mapsto$ 
${\cal X}_{\lambda}^{*}Q$ to a tensor ${\cal X}_{\lambda}^{*}Q$
on ${\cal M}_{0}$ and we expand
\begin{equation}
  \label{eq:Bruni-35}
  \left.{\cal X}^{*}_{\lambda}Q_{\lambda}\right|_{{\cal M}_{0}}
  =
  Q_{0}
  + \lambda {}^{(1)}_{\;\cal X}\!Q
  + \frac{1}{2} \lambda^{2} {}^{(2)}_{\;\cal X}\!Q
  + O(\lambda^{3}).
\end{equation}
This defines the first- and the second-order perturbations
${}^{(1)}_{\;\cal X}\!Q$ and ${}^{(2)}_{\;\cal X}\!Q$ of a
physical variable $Q_{\lambda}$ under the gauge choice 
${\cal X}_{\lambda}$.


When we have two different gauges ${\cal X}_{\lambda}$ and
${\cal Y}_{\lambda}$, Eq.~(\ref{eq:Bruni-35}) defines two
different representations of the $n$-th order perturbations
${}^{(n)}_{\;\cal X}\!Q$ and ${}^{(n)}_{\;\cal Y}\!Q$ on 
${\cal M}$, respectively.
We say that $Q$ is {\it gauge invariant up to order $n$} iff for
any two gauges ${\cal X}_{\lambda}$ and ${\cal Y}_{\lambda}$ the
following holds: ${}^{(k)}_{\;\cal X}\!Q$ $=$ 
${}^{(k)}_{\;\cal Y}\!Q$ for all $k$ with $k<n$.


The {\it gauge transformation} is simply the change of the point
identification map ${\cal X}_{\lambda}$ to another one.
If we have two different gauges ${\cal X}_{\lambda}$ and 
${\cal Y}_{\lambda}$, the change of the gauge choice from 
${\cal X}_{\lambda}$ to ${\cal Y}_{\lambda}$ is represented by
the diffeomorphism $\Phi_{\lambda}$ $:=$ 
$({\cal X}_{\lambda})^{-1}\circ{\cal Y}_{\lambda}$.
This diffeomorphism $\Phi_{\lambda}$ is the map $\Phi_{\lambda}$
$:$ ${\cal M}_{0}$ $\rightarrow$ ${\cal M}_{0}$ for each value
of $\lambda\in{\mathbb R}$.
Since the diffeomorphism $\Phi_{\lambda}$ does change the point
identification, the diffeomorphism $\Phi_{\lambda}$ is regarded
as the gauge transformation $\Phi_{\lambda}$ $:$ 
${\cal X}_{\lambda}$ $\rightarrow$ ${\cal Y}_{\lambda}$.


The diffeomorphism $\Phi_{\lambda}$ induces a pull-back from the
representation ${\cal X}_{\lambda}^{*}Q_{\lambda}$ in the gauge
${\cal X}_{\lambda}$ to the representation 
${\cal Y}_{\lambda}^{*}Q_{\lambda}$ in the gauge 
${\cal Y}_{\lambda}$, i.e., ${\cal Y}_{\lambda}^{*}Q_{\lambda}$
$=$ $\Phi^{*}_{\lambda} {\cal X}_{\lambda}^{*}Q_{\lambda}$.
Further, generic arguments of the Taylor expansion of the
pull-back of a tensor field on a manifold leads  
\begin{eqnarray}
  \Phi^{*}_{\lambda}{\cal X}_{\lambda}^{*}Q = {\cal X}_{\lambda}^{*}Q
  + \lambda {\pounds}_{\xi_{1}} {\cal X}_{\lambda}^{*}Q
  + \frac{\lambda^{2}}{2} \left\{
    {\pounds}_{\xi_{2}} + {\pounds}_{\xi_{1}}^{2}
  \right\} {\cal X}_{\lambda}^{*}Q
  + O(\lambda^{3}),
  \label{eq:Bruni-46-one} 
\end{eqnarray}
where $\xi_{1}^{a}$ and $\xi_{2}^{a}$ are the generators of the
diffeomorphism $\Phi_{\lambda}$.
The comparison with Eqs.~(\ref{eq:Bruni-35}) and
(\ref{eq:Bruni-46-one}) leads gauge transformation rule of each
order:
\begin{eqnarray}
  \label{eq:Bruni-47-one}
  {}^{(1)}_{\;{\cal Y}}\!Q - {}^{(1)}_{\;{\cal X}}\!Q = 
  {\pounds}_{\xi_{1}}Q_{0}, \quad
  {}^{(2)}_{\;\cal Y}\!Q - {}^{(2)}_{\;\cal X}\!Q = 
  2 {\pounds}_{\xi_{(1)}} {}^{(1)}_{\;\cal X}\!Q 
  +\left({\pounds}_{\xi_{(2)}}+{\pounds}_{\xi_{(1)}}^{2}\right)Q_{0}.
\end{eqnarray}


Inspecting the gauge transformation rules
(\ref{eq:Bruni-47-one}), we can define the gauge invariant
variables for perturbations for arbitrary matter
fields\cite{KNs}.
We expand the metric on ${\cal M}_{\lambda}$ is pulled back to
${\cal X}^{*}_{\lambda}\bar{g}_{ab}$ on ${\cal M}_{0}$ and it
expanded as ${\cal X}^{*}_{\lambda}\bar{g}_{ab}$ $=$ $g_{ab}$
$+$ $\lambda {}_{{\cal X}}\!h_{ab}$ $+$ 
$\frac{\lambda^{2}}{2} {}_{{\cal X}}\!l_{ab}$ $+$
$O^{3}(\lambda)$, where $g_{ab}$ is the metric on 
${\cal M}_{0}$.
First, we assume that {\it we already know the procedure for
  finding gauge invariant variables for the linear metric
  perturbations}, i.e, $h_{ab}$ is decomposed as $h_{ab}$ $=:$
${\cal H}_{ab}$ $+$ ${\pounds}_{X}g_{ab}$, where 
${}_{{\cal Y}}\!{\cal H}_{ab}$ $-$ 
${}_{{\cal X}}\!{\cal H}_{ab}$ $=$ $0$, and 
${}_{{\cal Y}}\!X^{a}$ $-$ ${}_{{\cal X}}\!X^{a}$ $=$
$\xi^{a}_{(1)}$.
This assumption is correct at least in the case of cosmological
perturbations\cite{KNs}.
Once we accept this assumption, we can show that the
second-order metric perturbation $l_{ab}$ is decomposed as
$l_{ab}$ $=:$ ${\cal L}_{ab}$ $+$ $2{\pounds}_{X}h_{ab}$ $+$
$\left( {\pounds}_{Y} - {\pounds}_{X}^{2} \right) g_{ab}$, where
${}_{{\cal Y}}\!{\cal L}_{ab}$ $-$ ${}_{{\cal X}}\!{\cal L}_{ab}$
$=$ $0$ and ${}_{{\cal Y}}\!Y^{a}$ $-$ ${}_{{\cal X}}\!Y^{a}$ $=$
$\xi_{(2)}^{a}$ $+$ $[\xi_{(1)},X]^{a}$.
Furthermore, using the first- and second-order gauge variant
parts, $X^{a}$ and $Y^{a}$, of the metric perturbations, the
gauge invariant variables for an arbitrary field $Q$ other than
the metric are given by  
\begin{eqnarray}
  {}^{(1)}\!{\cal Q} := {}^{(1)}\!Q - {\pounds}_{X}Q_{0}
  , \quad
  {}^{(2)}\!{\cal Q} := {}^{(2)}\!Q - 2{\pounds}_{X}{}^{(1)}\!Q 
  - \left({\pounds}_{Y}-{\pounds}_{X}^{2}\right)Q_{0}
  .
  \label{eq:matter-gauge-inv-def-1.0}
\end{eqnarray}
These imply that any first- and second-order perturbations is
always decomposed into gauge invariant and gauge variant parts
as Eqs.~(\ref{eq:matter-gauge-inv-def-1.0}), respectively.


\section{Conclusions}


We have shown the general procedure to find gauge-invariant
variables in the second-order general relativistic perturbation
theory through the precise treatments of ``gauges''.
We also showed that this general procedure is applicable to
cosmological perturbations and developed the second-order
cosmological perturbation theory in gauge invariant
manner\cite{KNs}.
Due to the general covariance in general relativity, all
equations are given in terms of gauge invariant variables.
We are planning to apply this second-order perturbation theory
to clarify the non-linear physics in Cosmic Microwave
Background\cite{WMAP}.


Besides the application to cosmology, we are also planning to
apply the above general framework to black hole perturbations,
perturbations of a star, and post-Minkowski description of a
binary system.
In conclusion, the definitions
(\ref{eq:matter-gauge-inv-def-1.0}) of gauge invariant variables
have very wide applications.


%


\section*{Acknowledgements}

This work is partially supported by Foundation for Promotion of
Astronomy.


\end{document}